%% file: main.tex
\documentclass[conference]{IEEEtran}

\usepackage[USenglish]{babel}	%
\usepackage[final]{graphicx}
	\graphicspath{{/}}
	\DeclareGraphicsExtensions{.pdf,.eps,.png,.jpg}
\usepackage{dblfloatfix}
\usepackage[caption=false]{subfig}
\usepackage{amsmath,amssymb}
\usepackage{bm}	%
\usepackage[binary-units,per-mode=reciprocal]{siunitx}
\usepackage{booktabs}
\usepackage{pgfplots}
	\pgfplotsset{%
		compat=1.15,
		table/col sep=comma,
	}
	\usepgfplotslibrary{colorbrewer}
\usepackage{pgfplotstable}
	\pgfplotstableset{col sep=comma}
	\pgfplotstableset{%
		results table/.style={%
			every head row/.style={%
				before row=\toprule,
				after row=\midrule,
			},
			every last row/.append style={after row=\bottomrule},
			every column/.append style={string type},
			assign column name/.style={/pgfplots/table/column name=\mbox{##1}},
			first column/.style={%
				column type/.add={@{}}{}
			},
			last column/.style={%
				column type/.add={}{@{}}
			},
		}
	}
\usepackage{makecell}
	
	\newcommand{\tableheaderx}[2]{\makecell{#1\\(\si{#2})}}

\usepackage{paralist}
\usepackage{boldmath-fix}
\usepackage[nolist,smaller]{eee-acronyms}
\usepackage{acronym-patch,acronym-abstract,acronym-cap}
\usepackage{acronym-alwayslong}
	\acronymalwayslong{HDC}
	\acronymalwayslong{LA}
	\acronymalwayslong{MEP}
	\acronymalwayslong{MPPT}
	\acronymalwayslong{STA}
	\acronymalwayslong{TA}
	\acronymalwaysshort{RTL}
\usepackage{eee-circuit,eee-macros}
\usepackage{boolean}
\input{report-process}

\providecommand{\tikzinput}[1]{\includegraphics{tikzfigure-#1}}%

\input{macros}	%
\let\oldgate\gate
\renewcommand{\gate}[1]{\oldgate{#1} gate}
\usepackage{suffix}
\WithSuffix\newcommand\gate*{\oldgate}
\newcommand*\gates[1]{\gate{#1}s}
\newcommand*\nth[1]{$#1^\mathrm{th}$}

\newcommand*{\bname}[2][]{$\textsc{#2}_{#1}$}     %

\newcolumntype{L}{>{$}l<{$}}%
\newcolumntype{R}{>{$}r<{$}}%
\newcolumntype{C}{>{$}c<{$}}%

\usepackage[noabbrev,capitalise]{cleveref} %
	\crefname{equation}{}{}	%
	\crefname{assumption}{Requirement}{Requirements}
	\crefname{method}{Method}{Methods}

\begin{document}
	\hyphenation{sub-threshold super-threshold}
	\title{Low-Latency Asynchronous Logic Design\\for Inference at the Edge}
	\author{%
		\IEEEauthorblockN{Adrian~Wheeldon, Alex~Yakovlev, Rishad~Shafik and Jordan~Morris}
		\IEEEauthorblockA{%
			Microsystems Group, Newcastle University, Newcastle Upon Tyne, UK\\
			Email: adrian.wheeldon@ncl.ac.uk
		}
	}
	\maketitle

	\input{abstract}
	\input{intro}
	\input{tm}
	\input{method}

	\input{infer}
	\input{conclusion}

	\bibliography{}

\end{document}

%% file: report-process.tex
\csname @namedef\endcsname{processlibname@faraday90ll}{\pnoun{Faraday \SI{90}{\nano\metre} LL}}
\csname @namedef\endcsname{processlibname@faraday65ll}{\pnoun{Faraday \SI{65}{\nano\metre} LL}}
\csname @namedef\endcsname{processlibname@pipistrelle4}{\pnoun{Full Diffusion}}
\csname @namedef\endcsname{processlibname@umc65ll}{\pnoun{UMC LL}}
\def\processlibname#1{%
	\expandafter\ifx\csname processlibname@#1\endcsname\relax
		\expandafter\errmessage{Library name `#1' didn't match any that I know.}\errorstopmode
	\else
		\csname processlibname@#1\endcsname%
	\fi
}
\csname @namedef\endcsname{processlibnameshort@faraday90ll}{\pnoun{F\num{90} LL}}
\csname @namedef\endcsname{processlibnameshort@faraday65ll}{\pnoun{F\num{65} LL}}
\csname @namedef\endcsname{processlibnameshort@pipistrelle4}{\pnoun{FD}}
\csname @namedef\endcsname{processlibnameshort@umc65ll}{\pnoun{UMC LL}}
\def\processlibnameshort#1{%
	\expandafter\ifx\csname processlibnameshort@#1\endcsname\relax
		\expandafter\errmessage{Library name `#1' didn't match any that I know.}\errorstopmode
	\else
		\csname processlibnameshort@#1\endcsname%
	\fi
}

%% file: macros.tex
\newcommand{\pnoun}[1]{\textsc{#1}}
\newcommand*{\signal}[2][]{\ensuremath{\mathsf{#2}_\mathsf{#1}}}
\newcommand*{\nsignal}[2][]{\ensuremath{\bnot{\mathsf{#2}_\mathsf{#1}}}}
\newcommand*{\prail}[2][]{\ensuremath{\signal[#1]{#2}^\mathsf{p}}}
\newcommand*{\nrail}[2][]{\ensuremath{\signal[#1]{#2}^\mathsf{n}}}
\let\oldvec\vec
\renewcommand*{\vec}[2][]{\ensuremath{\oldvec{\mathsf{#2}}_{#1}}}

\newcommand{\cel}{C-element}
\newcommand{\cels}{C-elements}

\newcommand{\sr}{single-rail}
\newcommand{\Sr}{Single-rail}

\newcommand{\dr}{dual-rail}
\newcommand{\Dr}{Dual-rail}

\newcommand{\popcount}{\popcounta}

\newcommand{\PopCount}{\PopCounta}

\newcommand{\popcounta}{population count}

\newcommand{\PopCounta}{Population Count}

\newcommand*{\feat}[1]{f}
\newcommand*{\finput}{feature input}

\newcommand*{\finputs}{\finput{}s}

\newcommand*\tspcw{\ensuremath{t_{\sptocw}}}
\newcommand*\tcwsp{\ensuremath{t_{\cwtosp}}}

\renewcommand{\vdd}[1]{%
        \def\temp{#1}\ifx\temp\empty
                \ensuremath{V_{\mathrm{\textsc{dd}}}}%
        \else
                \ensuremath{V_{\mathrm{\textsc{dd},#1}}}%
        \fi
}

\newcommand{\trans}{\texorpdfstring{\ignorespaces\,$\to$\,\ignorespaces}{ to }}
\newcommand{\codeword}{valid}

\newcommand{\spacer}{spacer}

\newcommand{\sptocw}{\textsc{s\trans{}v}}

\newcommand{\cwtosp}{\textsc{v\trans{}s}}
\newcommand{\Cwtosp}{\textsc{V\trans{}s}}

%% file: abstract.tex
\begin{abstract}
	Modern \acf{IoT} devices leverage \ac{ML} inference using sensed data
	on-device rather than offloading them to the cloud. Commonly known as
	\emph{inference at-the-edge}, this gives many benefits to the users,
	including personalization and security. However, such applications
	demand high energy efficiency and robustness.
	In this paper we propose a method for reduced area and power overhead of
	self-timed early-propagative asynchronous inference circuits, designed
	using the principles of \acp{LA}. Due to natural resilience to timing as
	well as logic underpinning, the circuits are tolerant to variations in
	environment and supply voltage whilst enabling the lowest possible
	latency. Our method is exemplified through an inference datapath for a
	low power \ac{ML} application. The circuit builds on the
	\ac{TM} algorithm further enhancing its energy efficiency.
	Average latency of the proposed circuit is reduced by $10\times$ compared with
	the synchronous implementation whilst maintaining similar
	area.
	Robustness of the proposed circuit is proven through post-synthesis
	simulation with \SIrange{0.25}{1.2}{\volt} supply. Functional
	correctness is maintained and latency scales with gate delay as voltage
	is decreased.
\end{abstract}

%% file: intro.tex
\section{Introduction}

There is an accelerating demand for connected devices in the
\ac{IoT}~\cite{Capra2019}. Such devices often comprise a sensing aspect,
collecting environmental or personal data, for providing useful monitoring and decisions for transforming our everyday life. The
sensors collect vast amounts of data which must be processed into a usable or
more manageable form. Traditionally this was done by offloading the data into
cloud compute servers, usually over a wireless medium. However this paradigm is
quickly becoming unmaintainable as \ac{IoT} devices expand well into the
billions~\cite{shafik2018real}. Generated data sizes become overwhelming, wireless data transmissions
violate power budgets, and we see a shift towards data processing at the
edge~\cite{arm-trillium}.
Designers of \ac{IoT} products are turning to \ac{ML} in order to extract
meaningful features from the sensed data. Such products are often powered by
batteries or energy harvesters which demand low power and energy efficiency, as
well as robustness to supply variations~\cite{shafik2018real}.

There are several \ac{ML} algorithms which may be suited to such applications,
with \acp{NN} in widespread usage thanks to their often state-of-the-art
accuracy and powerful hardware\slash software ecosystem. \Ac{HDC} has also emerged in recent years with applications in low
power systems~\cite{Burrello2020}. Recently the \ac{TM} algorithm has been proposed as a
promising \ac{ML} algorithm based on \acp{TA}---specialized \acp{LA}. The
\acp{TA} use reinforcement learning locally, together creating an ensemble
learning effect on the global scale which is used to compose logic clauses.
Existing hardware based on \ac{TM} offers a new direction for \ac{ML} whose
inference engine is based on logic with little arithmetic~\cite{Wheeldon2020b}.
The logic-based underpinnings of the \ac{TM} algorithm provide opportunities
for low power and energy efficient \ac{ML} hardware design in the \ac{IoT}.

In this work we apply an asynchronous circuit design
methodology~\cite{Wheeldon2019} to the \ac{TM} algorithm. By removing the pairing between clock and supply voltage as in the synchronous digital designs, it enables an aggressive
voltage scaling~\cite{Diamant2015} for reduced energy per inference and also
adds robustness to environmental variations. Although we use \ac{ML} as the key
application driver, it is possible these techniques can also be applied in other
application areas. Our method is built on \emph{\dr{}} circuits with early
propagation~\cite{Brej2006}. \Dr{} is an asynchronous circuit design style in the family of
\ac{QDI} circuits. It is inherently robust to circuit delay variations which
means it can operate across a wide range of supply voltages and temperatures.
This usually comes at the cost of duplicated logic and \ac{CD}
overhead~\cite{Yakovlev2013}. In our design we carefully select circuit topology
to minimize such duplications. Additionally we use timing optimizations to reduce
overhead from \ac{CD}.

\subsubsection*{Nomenclature}
Positive- and negative-rail signals are denoted \prail{x} and \nrail{x}
respectively.
\signal[m]{x} denotes the \nth{m} signal in the bit vector \signal{x}.
\Cwtosp{} denotes a transition on a \dr{} signal from a valid
codeword to a spacer. Vice versa for \sptocw{}.

\subsubsection*{Major Contributions of this paper}
\begin{enumerate}
        \item application of early-propagative, reduced-overhead self-timed
                \dr{} circuits to \ac{ML} inference; and
        \item analysis of operand and delay probability distributions in the
                \ac{ML} inference circuit.
\end{enumerate}

\subsubsection*{Paper Organization}
\Cref{sec:tm} introduces the concepts of the \ac{TM} algorithm.
\Cref{sec:method} first briefly introduces the principles of \dr{} circuits
before describing our reduced \ac{CD} scheme. \Cref{sec:infer} presents our
\dr{} inference datapath design with in-depth analyses. We finally conclude our
findings in \cref{sec:conclusion}.

%% file: tm.tex
\section{Tsetlin Machine Overview}\label{sec:tm}

\begin{figure}
    \centering
    \tikzinput{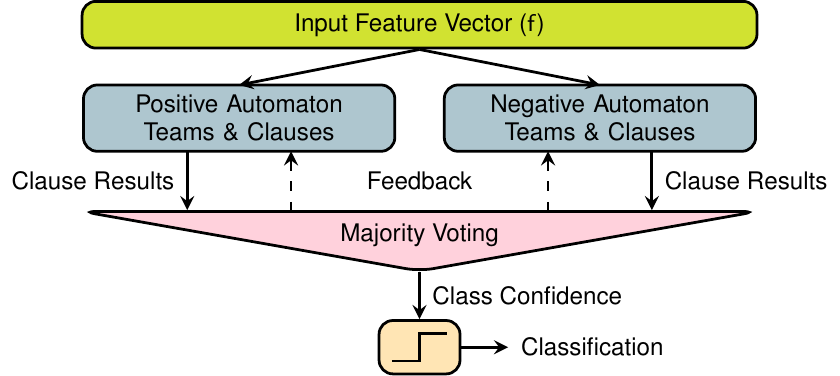}
    \caption{%
        Simplified overview of a single \acf{TM} classifier.
    }\label{fig:tsetlin}
\end{figure}

The main inference component of the \ac{TM} is the conjunctive clause which uses
propositional logic expressions to produce a vote. The composition of each
clause (determined by inclusion of literals) is controlled by the action outputs of a team of
\aclp{TA}.
For inference, the \acp{TA} themselves are not required.
Following a number of reinforcement steps, the automata decide whether their associated literal should be excluded from (action 1) or included in (action 2) the clause.
\Cref{fig:tsetlin} illustrates a \ac{TM} classifier with automaton teams and
conjunctive clauses as one block for brevity.

Each clause
can produce a vote for its class. Half of the clauses can vote
positively, while the other half of the clauses can vote negatively. The inclusion of
inhibition in the voting system enables non-linearity in the inference
process. The votes are summed in a majority vote to produce a collective result
which gives an indication of confidence. This confidence is used to influence future decisions of the automata~\cite{Granmo2018}.

A simple thresholding function can be used
to generate the final classification output. If the votes are positive (or
zero), the input data is determined to belong to the class. For a negative sum
the input data is determined to be \emph{not} in the class.

For purposes of studying inference, we abstract the \ac{TA} action outputs
to the circuit's environment and concentrate on only the clauses
calculation and majority voting.

%% file: method.tex
\section{Self-timing Methodology}\label{sec:method}

In \dr{} logic two wires are used to encode a codeword. For a single bit
$\signal{x}$, the \dr{} encoding consists of the positive and negative rails
$\{\prail{x},\nrail{x}\}$. $\signal{x}=0$ is encoded as $\{0,1\}$, and
$\signal{x}=1$ is encoded as $\{1,0\}$. One of the remaining states---$\{0,0\}$
or $\{1,1\}$---is chosen to represent the empty state, referred to as a
\emph{spacer}, which separates valid codewords temporally so they can be
distinguished from each other. Care must be taken to correctly handle \spacer{}
in the design, otherwise data hazards could occur where one \codeword{}
overtakes another~\cite{Sokolov2006}. The remaining state is forbidden and must
be avoided by design.

In our design we abide by the following requirements to ensure correct circuit
operation:
\begin{enumerate}
	\item Monotonic switching at the \acsp{PI}.\label[assumption]{ass:monotonic-pi}
	\item Monotonic switching within the circuit.\label[assumption]{ass:monotonic-circuit}
	\item Acknowledgment of \sptocw{} on \acsp{PO}.\label[assumption]{ass:cd}
	\item \Cwtosp{} on \acsp{PO} and internal signals before new
		\acsp{PI} applied.\label[assumption]{ass:timing}
	\item \acsp{PI} must transition \sptocw{} and \cwtosp{} for each operand.\label[assumption]{ass:determinism}
	\item \acsp{PI} transition \cwtosp{} only after \sptocw{} on \acsp{PO}.\label[assumption]{ass:pi-cwtosp}
\end{enumerate}

\Cref{ass:monotonic-pi,ass:pi-cwtosp,ass:determinism} are assumed as part of the
circuit's environment. To ensure \cref{ass:monotonic-circuit}, the circuit must
be constructed solely from unate logic gates. To maintain monotonicity we must
exclude non-unate logic gates (\eg{} \gate*{xor} and \gate*{xnor}) from our
library when generating the \dr{} netlist.
\Cref{ass:cd} is taken care of by \ac{CD} insertion.
\Cref{ass:timing} can either be assumed as part of the environment or a delay
can be added to the falling edge of \ac{CD} assertion. The latter will be
discussed in \cref{sec:timing-assumption}.

\subsection{Reduced Completion Detection Scheme}\label{sec:timing-assumption}

\ac{CD} which acknowledges both \sptocw{} and \cwtosp{} at the \acp{PO} is expensive to implement due to the
vast amount of complex \cels{} required~\cite{Sparso2001}. By indicating only \sptocw{}
transitions we can significantly reduce the overhead of \ac{CD} by using a small
number of simple gates.

Full \ac{CD} on \emph{internal signals} is even more costly and
removes the possiblity of early propagation. Its job is to ensure \sptocw{} and \cwtosp{} occurs on internal nets for each operand. Internal \ac{CD} can be safely
omitted by giving a grace period for the internal signal to reset to spacer
before applying new \acp{PI}. Codeword validity and correct operation can still
be guaranteed as long as \cref{ass:pi-cwtosp,,ass:determinism,ass:timing} are
met.

In order to meet \cref{ass:timing} there must be a sufficient grace period from
application of \spacer{} at the \acp{PI} until application of the next
\codeword{} at the \acp{PI}. The grace period can be determined by using \ac{STA}
to find the maximum possible \sptocw{} time on all nodes of the circuit.
Consequently the grace period can be guaranteed by either
\begin{inparaenum}
	\item the circuit environment waiting for the required grace period; or
	\item an appropriate delay built into the done signal of the
		\ac{CD}.\label[method]{meth:delay}
\end{inparaenum}
The required delay can be calculated as
\newcommand*\td{\ensuremath{t_\mathrm{d}}}%
\newcommand*\tnet{\ensuremath{t_\mathrm{int}}}%
\newcommand*\tio{\ensuremath{t_\mathrm{io}}}%
\newcommand*\tdone{\ensuremath{t_\mathrm{done}}}%
$ \td = \tnet - \tio $,
where \tnet{} is the maximum internal net \cwtosp{} time, and \tio{}
is the maximum \cwtosp{} time from the \acp{PI} to \acp{PO}.
\tnet{} must include false paths. It is these false path which lead to the
distinction between \tnet{} and \tio{}. Since there may be some margin added to
\td{}, or due to implementation of the delay \td{} may be greater than the
requirement, the actual timing of the 1\trans{}0 transition of \signal{done} can be calculated by
$ \tdone{}_{1\rightarrow{}0} = \tio + \td $.

%% file: infer.tex
\section{Inference Datapath}\label{sec:infer}

The inference datapath of the \ac{TM} is derived
from the full \ac{TM} diagram (\cref{fig:tsetlin}). The \acp{TA} and their feedback
are not required for inference.
Only the \emph{exclude} action output is required from the \acp{TA} teams. In the
diagram this is abstracted to the \ac{PI} \signal{e}.
We split the majority voting of the \ac{TM} into two sections. Firstly we
distinctly count all positive votes and negative votes by means of \popcount{}s.
Secondly the two counts are compared using a magnitude comparator to determine
the winner. The result of the comparison is taken as the classifier outcome.

There are several ways to construct the circuit architecture. This architecture
has been chosen due to the simplicity and efficiency of the asynchronous
magnitude comparator as will become clear in \cref{sec:comparator}.

All \acp{PI} and \acp{PO} of the circuit are \dr{} encoded. These can interface
natively with other \dr{} signals, or with synchronous circuits using
converters~\cite{Sokolov2006}.

\subsection{Clause Calculation}

The \signal{e} input to the inference datapath controls whether the
corresponding \finput{} (\signal{f}) will be excluded from a clause computation.
We use \gates{or} to form a mask of each \finput{} in each clause.
The partial clause
values, \signal{pc}, must be aggregated using an \gate*{and} tree in order
to evaluate the entire clause comprising input from all \signal{f} and their
associated automaton actions. The \emph{exclude} signals (\signal{e}) from the
\acp{TA} mask \signal[m]{f} and \nsignal[m]{f} \finputs{} causing logic-1 at the
\gate{and} inputs. If \signal[2m]{e} (\resp{} \signal[2m+1]{e}) is logic-0
(\ie{} the \finput{} should be \emph{included} in the clause calculation), the
value of \signal[m]{f} (\resp{} \nsignal[m]{f}) is passed through to the
\gate{and} to be evaluated.
The partial clause evaluation circuit is replicated as many times as there are
\finputs{} to the \ac{TM}.

Since \signal[m]{f} will be \dr{} encoded in our system, we do not need to
generate \nsignal[m]{f} internally. By performing direct mapping of a \sr{}
circuit, and along with negative gate optimization~\cite{Sokolov2006}, we arrive
at the optimized \dr{} circuit in \cref{fig:clause-dr-opt2}. All signal paths in
this circuit have a single inversion---satisfying spacer requirements and giving
the block an \emph{inverting spacer} overall.

\subsection{\PopCount{}}

\begin{figure}[h]
	\tikzinput{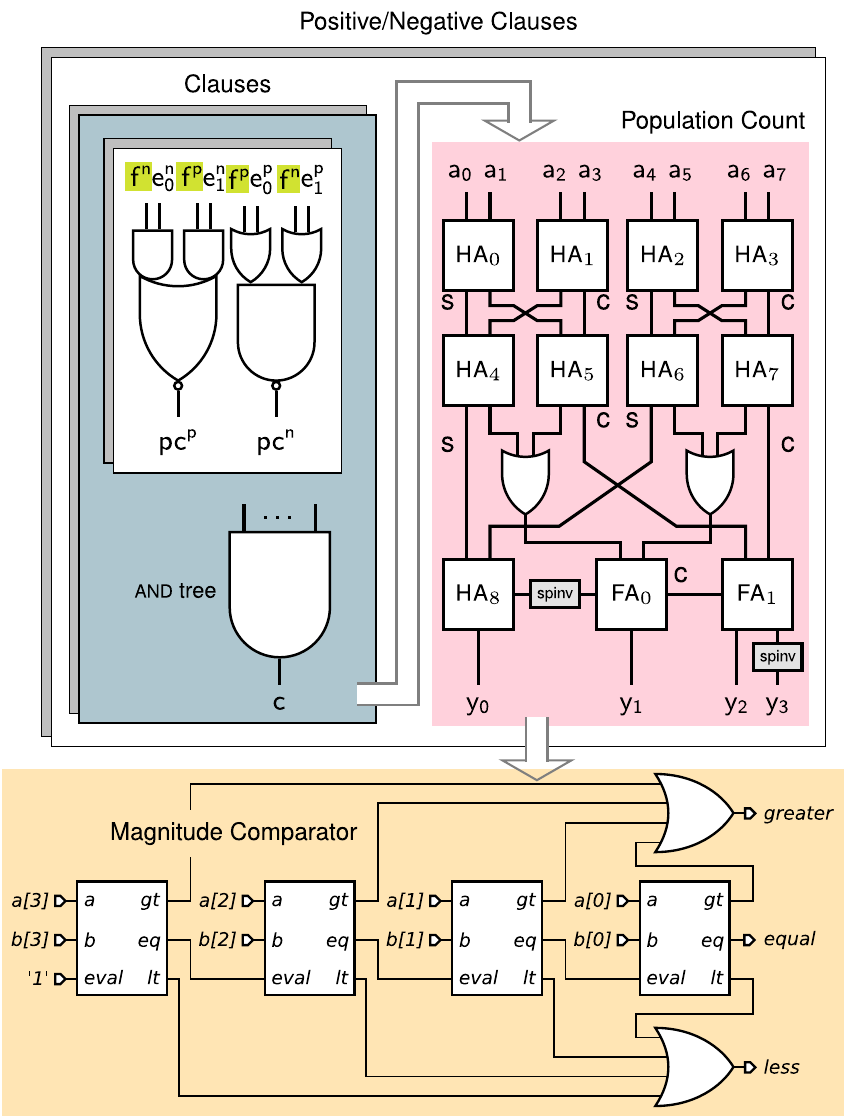}
	\caption{%
		Block diagram of the \acf{TM} inference path. Including \dr{}
		circuits for partial clause evaluation; and \popcount{}
		architecture, where each wire represents two signals with \dr{}
		encoding.
		spinv: spacer inverter.
	}\label{fig:dr-circuits}\label{fig:clause-dr-opt2}\label{fig:popcount-sr}
\end{figure}

We base our \popcount{} circuit on the optimized design of
Dalalah~\cite{Dalalah2006}. The eight-input design comprises nine half-adders,
two full-adders and two \gates{or} and is illustrated in \cref{fig:popcount-sr}.
Each wire in the diagram represents two signals which form the \dr{} encoding.
The \dr{} \gate{or} is internally constructed from one \gate{or} and one
\gate{and}. The \dr{} half-adders are constructed using two complex gates and
two simple gates each. There is no spacer inversion within the half-adders as
all signal paths have an even number of inversions. The \dr{} full-adder is
constructed from six complex gates, two simple gates and four
inverters~\cite{Wheeldon2019}. It has inverted spacers on carry-in and carry-out
with respect to the other inputs and outputs, therefore we must accommodate for
these in the \popcount{} design by adding spacer inverters:
\begin{inparaenum}
	\item between \bname[8]{ha} and \bname[0]{fa};
	\item between \bname[1]{fa} and the \signal[3]{y} output.
\end{inparaenum}
The resulting \dr{} \popcount{} circuit has no spacer inversion overall,
therefore the output spacer will have the same polarity as the input spacer.

\subsection{Magnitude Comparator}\label{sec:comparator}

The magnitude comparator compares the number of votes from the positive and
negative \ac{TA} teams. A larger number of positive votes indicates that the
input pattern belongs to the class in question, and conversely, a larger number
of negative votes indicates that the input pattern \emph{does not} belong to the
class in question.

The magnitude comparator is based on a request architecture~\cite{Wheeldon2019}
and compares the operands in bit-pairs, starting from the most significant bit.
Once a difference is found, the answer is known, and the remainder of the bits
need not be compared. This architecture enables huge average-case latency
improvement over a synchronous counterpart. Energy savings are also made by
due to saved switching power on the lower bits when the operands differ by a large
magnitude.

Since the comparator's outputs (less, equal, and greater) are mutually exclusive,
we take advantage of this in the asynchronous design. We use a
1-of-3 encoding on the output instead of the usual \dr{}---1-of-n encoding being
a superset of \dr{}. Provided a \spacer{} seperates the \codeword{}s, the
switching of 1-of-n codes is monotonic~\cite{Bainbridge2003},
therefore satisfying \cref{ass:monotonic-circuit}.
Without this trick, three sets of \dr{} signals would be required at the comparator outputa at the expense of more logic to drive these signals.
The inputs to the comparator are \dr{} encoded.

\subsection{Inference Datapath Results}

\begin{table*}[!htb]
	\centering
	\caption{%
		Comparison of \sr{} and \dr{} circuits after synthesis.
	}\label{tab:results}
	\input{results-table}
\end{table*}

The inference datapath was synthesized using \pnoun{Synopsys Design Compiler}
for two different \SI{65}{\nano\metre} silicon libraries. \processlibname{umc65ll} is a
commercially available, low-leakage library which we use with nominal
\SI{1.2}{\volt} supply and TT corner. \processlibname{pipistrelle4} is a custom
library aimed at high performance subthreshold
operation~\cite{Morris2017}. It uses a full diffusion sizing strategy with
non-minimum-length transistors in order to mitigate subthreshold effects. For
this silicon library the circuit is first synthesized at TT corner for nominal
\SI{1.2}{\volt} supply and results are shown with supply voltage in the range
\SIrange{0.25}{1.2}{\volt}.

Results in \cref{tab:results} show similar cell areas for both \sr{} and \dr{}
designs for each silicon library. This is possible due to the careful choice
of \dr{} circuit architecture and the reduced completion detection scheme.
The \dr{} clause computation and magnitude comparator are more area efficient
than their \sr{} conterparts due to exploitation of \dr{} encoding and clever
use of 1-of-3 encoding respectively.

For
the area of the sequential cells we count flip-flop area for the \sr{} designs
and \cel{} area for \dr{} designs.
The sequential area is similar between designs, despite the \dr{} design
having twice as many sequential cells due to the doubled input rails. The
\dr{} circuit uses \cels{} as latches. These comprise four simple gates in the
\processlibname{pipistrelle4} library (due to lack of \gate*{AOI32} cells) and a single complex gate
in the \processlibname{umc65ll} library. Note that the cell area
varies dramatically between the libraries due to transistor sizing%
---\processlibname{umc65ll} being minimally-sized for superthreshold
and \processlibname{pipistrelle4} larger for subthreshold operation. The
\emph{number} of cells does not vary significantly.

Latency is measured from \sptocw{} in the \dr{} designs, and the clock period
defines the latency for \sr{} designs. The \dr{} circuit enables $10\times$ reduction in
average latency thanks to early propagation. Average throughput is worsened
however, due to the lengthened logic path and the need for the additional
\cwtosp{} transition. Although the \dr{} switching power is greater due to
higher inherent activity factor, the computation energy is reduced due to
increased throughput.

Throughput period is defined by the \sr{} circuit's clock period. For the \dr{}
design, throughput period is determined by $\tspcw + \tspcw$ so that
the \acp{PI} are ready for the next operand. \tcwsp{} has the same magnitude as
$\max (\tspcw)$.

\begin{figure}
	\centering
	\tikzinput{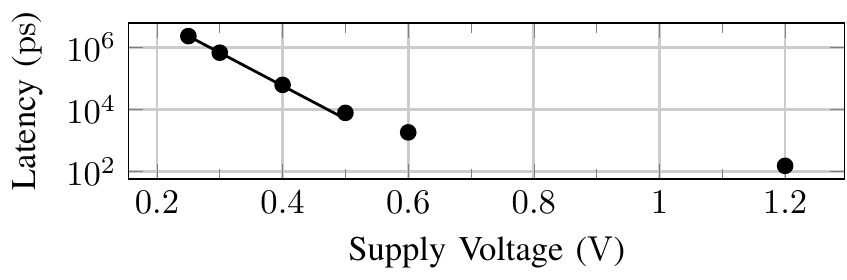}
	\caption{%
		Scaling of \dr{} datapath latency with supply voltage for the
		\processlibname{pipistrelle4} library.
	}\label{fig:infer-avg-pipi4}
\end{figure}

\Cref{fig:infer-avg-pipi4} shows the effects of supply voltage on datapath
latency. The latency increases exponentially as the supply voltage is reduced
from \SIrange{0.6}{0.25}{\volt}. The key point is that the circuit functionality
is guaranteed across the whole supply voltage range thanks to the requirements
in \cref{sec:method} and without any alteration to the hardware.

%% file: results-table.tex
\begingroup
\sisetup{%
	round-mode=figures,
	round-precision=2,
	scientific-notation=fixed,
	fixed-exponent=0,
}
\setlength{\tabcolsep}{0.5em}

\pgfplotstableread{tminfer.stats}\tabletminfer
\pgfplotstablecreatecol[
	expr={1/(\thisrow{avg-optime} + \thisrow{max-spacertime})}
]{avg-infer}{\tabletminfer}

\pgfplotstabletypeset[
	create on use/tech/.style={create col/set list={\processlibname{umc65ll},,\processlibname{pipistrelle4},}},
	results table,
	columns={tech,design,area,seq-area,avg-power,leakage-power,avg-optime,
		max-optime,max-spacertime,avg-infer},
	columns/tech/.style={%
		column name=Technology,
		column type={l},
		first column,
	},
	columns/design/.style={%
		column name=Design,
		column type={l},
		string replace={tminfer-sr}{\Sr{}},
		string replace={tminfer-burst}{Proposed \Dr{}},
	},
	columns/area/.style={%
		column name=\textbf{Cell Area},
		column type={S[table-format=4]},
	},
	columns/seq-area/.style={%
		column name=\makecell{Sequential\\Area},
		column type={S[table-format=4]},
	},
	columns/avg-power/.style={%
		column name=\tableheaderx{%
			Avg.\@ Power
		}{\micro\watt},
		column type={S[table-format=4]},
		multiply with=1e9,
	},
	columns/leakage-power/.style={%
		column name=\tableheaderx{Leakage Power}{\nano\watt},
		column type={S[table-format=4]},
		multiply with=1e12,
	},
	columns/avg-optime/.style={%
		column name=\textbf{\tableheaderx{Avg.\@ Latency}{\pico\second}},
		column type={S[table-format=4]},
	},
	columns/max-optime/.style={%
		column name=\tableheaderx{Max Latency}{\pico\second},
		column type={S[table-format=4]},
	},
	columns/max-spacertime/.style={%
		column name=\tableheaderx{\tcwsp}{\pico\second},
		column type={S[table-format=3]},
		string replace={0}{\textemdash},
	},
	columns/avg-infer/.style={%
		column name=\makecell{Avg.\@ Inferences\\(Millions \si{\per\second})},
		column type={S[table-format=4]},
		multiply with=1e6,
		last column,
	},
]{\tabletminfer}
\endgroup

%% file: conclusion.tex
\section{Conclusion}\label{sec:conclusion}

In this paper we have demonstrated an asynchronous, self-timed
inference datapath design with area and power of equal orders of magnitude to the
synchronous equivalent. Early propagation enables $10\times$ lower inference latency
than the equivalent synchronous circuit on average. The savings
are enabled by a reduced \ac{CD} scheme which can be applied to any \dr{}
asynchronous circuit. The new scheme introduces a timing assumption which can be
incorporated into the \ac{CD} circuit, so that the circuit environment does not need to be adapted.
This type of low-latency circuit can have applications in speech
recognition for wearables and other low-power applications where inference
latency is of particular importance.

In future work we will apply asynchronous design styles to the training
datapath of the \ac{TM} algorithm in order to enable a fully-asynchronous
\ac{ML} hardware capable of on-chip learning.